\documentclass[prl,twocolumn,amsmath,amssymb,showpacs,floatfix]{revtex4}
\usepackage[latin9]{inputenc}
\usepackage{graphicx,bm,setspace}
\usepackage{color}
\usepackage{graphicx}
\makeatletter
\def\graphicscale{\twocolumn@sw{0.33}{0.4}}
\makeatother

\def\spose#1{\hbox to 0pt{#1\hss}}
\def\lesssim{\mathrel{\spose{\lower 3pt\hbox{$\mathchar"218$}}
 \raise 2.0pt\hbox{$\mathchar"13C$}}}
\def\gtrsim{\mathrel{\spose{\lower 3pt\hbox{$\mathchar"218$}}
 \raise 2.0pt\hbox{$\mathchar"13E$}}}

\def\<{\langle}
\def\>{\rangle}

\newcommand*{\beq}{\begin{eqnarray}}
\newcommand*{\eeq}{\end{eqnarray}}
\newcommand*{\bea}{\begin{eqnarray}}
\newcommand*{\eea}{\end{eqnarray}}

\def\simge{\mathrel{%
       \rlap{\raise 0.511ex \hbox{$>$}}{\lower 0.511ex \hbox{$\sim$}}}}
\def\simle{\mathrel{
       \rlap{\raise 0.511ex \hbox{$<$}}{\lower 0.511ex \hbox{$\sim$}}}}

\makeatother

\begin{document}

\title{Field-controlled  columnar and planar patterning of cholesteric
colloids}

\author{G. D'Adamo$^1$, D. Marenduzzo$^2$, C. Micheletti$^1$,  E. Orlandini$^3$,}

\affiliation{$^1$ SISSA, International School for Advanced Studies, via Bonomea 265, I-34136 Trieste, Italy \\
$^2$SUPA, School of Physics and Astronomy, University of Edinburgh, Mayfield Road, Edinburgh EH9 3JZ, UK\\
$^3$ Dipartimento di Fisica e Astronomia, Universita' di Padova, Via Marzolo 8, 35131 Padova, Italy}

\begin{abstract}
We study how dispersions of colloidal particles in a cholesteric liquid crystal behave under a
time-dependent electric field.
By controlling the amplitude and shape of the applied field wave, we show that the system can be
reproducibly driven out of equilibrium through different kinetic pathways and
navigated through a glassy-like free energy landscape encompassing many competing metastable
equilibria. Such states range from simple Saturn rings to complex structures
featuring amorphous defect networks, or stacks of disclination loops. A non-equilibrium electric field can also trigger the alignment of particles into
columnar arrays, through defect-mediated force impulses, or their
repositioning within a plane. Our results are promising in terms of providing
new avenues towards controlled patterning and self-assembly of soft colloid-liquid crystal composite materials.
\end{abstract}


\maketitle
Liquid crystal colloids are attracting considerable attentions for their potential as
novel and versatile materials. These include photonic crystals, self-quenched glasses, meta-materials, new biosensors and multistable devices~\cite{Ravnik-Photonic,Lavrentovich2011,Porenta2014topological,kang2001electro,Wood2011self,lin2011endotoxin,muvsevivc2006two,foffano2014dynamics}.
At the same time these composite materials are of considerable fundamental interest. In fact, colloidal
objects disrupt the orientational order of the hosting liquid crystal generating a remarkable array
of topological excitations, including simple Saturn rings~\cite{Poulin1997}, nematic
braids~\cite{ravnik2007entangled,vcopar2011nematic}, and exotic knotted or linked disclination
networks~\cite{tkalec2011reconfigurable,jampani2011colloidal,martinez2014mutually,machon2013knots}. While these structures may form with spherical particles in nematics or cholesterics, complex defect
patterns can also be achieved by dispersing colloidal particles with nontrivial topologies such as
handle-bodies, ribbons or knots~\cite{senyuk2013topological, liu2013nematic, irvine2014liquid,
campbell2014topological}.

Intriguingly, these defect textures can, in turn, mediate interactions between colloidal particles~\cite{Poulin1997,Stark2001,Yada2004,Tomar2012}, and create an inherently glassy free energy landscape. The associated elastic
energy barriers typically dwarf thermal noise and hence trap the system
in a variety of different metastable states. This property is key to
applications, because it opens the possibility to build novel devices capable of
switching between different soft metastable states, each having specific
optical or mechanical properties. Because the switchable states are long-lived,
these devices are multistable and exhibit memory effects, which are desirable features
for the design of new-generation energy-saving technologies such as
e-paper~\cite{yang1997bistable,tiribocchi2011bistable,araki2011memory,serra2011emergence,stratford2014self}.
Accordingly, devising novel colloid-liquid crystal composites with defect
properties that can be tuned by external intervention, represents a major
standing challenge of soft matter physics.

Towards this goal, here we explore a new way to create robust distinct defect
and particle patterns  by exploiting a time-dependent
electric field acting on a cholesteric hosting {electrically-neutral colloids.
As shown by the studies in refs.~\cite{PhysRevLett.87.165503,PhysRevLett.106.047801}, time-dependent electric fields hold great potential for being used as
powerful manipulation tools alongside with well-established techniques based on constant fields or optical forces~\cite{muvsevivc2004laser,lavrentovich2014transport}}.
The behaviour resulting from the competition between the nonequilibrium driving
mechanism and the free energy landscape complexity is very rich: even for a
single colloid, the time-dependent field can create a variety of defect
structures which depend sensitively on the details of the
external force. When several particles are dispersed in the cholesteric, this
external tunability can be harnessed to drive a reproducible global
rearrangement of the colloids from random to columnar or planar arrays. The
results can have practical implications for the realization of composites with
controllable optical, rheological or mechanical properties.


The system that we consider consists of a set of spherical colloids
embedded in a cholesteric liquid crystal, and subject to the action of a spatially-uniform,
but time-dependent, electric field, ${\mathbf E}$.
Accordingly, the Landau-de Gennes free energy density, $f$, results from the linear
superposition of the bulk, elastic-distortion and electric field terms, $f=f_{b}+f_{el}+f_E$.
Adopting the repeated indices summation convention, their expressions are~\cite{wright1989crystalline}:
\begin{eqnarray}
f_{b}&=&\tfrac{A_0}{2}\left(1-\tfrac{\gamma}{3}\right) Q^2_{\alpha\beta}-\tfrac{A_0\gamma}{3}Q_{\alpha,\beta}Q_{\beta\gamma}Q_{\gamma\alpha}\nonumber +\tfrac{A_0\gamma}{4}\left( Q^2_{\alpha\beta} \right)^2, \nonumber \\
f_{el}&=&\tfrac{L}{2}\left[(\partial_\beta Q_{\alpha,\beta })^2+(\epsilon_{\alpha\gamma\delta}\partial_\gamma Q_{\delta\beta}+2 q_0 Q_{\alpha\beta}  )^2 \right], \nonumber \\
f_{E}&=&-\tfrac{1}{12\pi}\epsilon_a E_\alpha Q_{\alpha\beta}E_\beta \,
\end{eqnarray}
\noindent where ${\bf Q}$ is the traceless and symmetric tensor order
parameter~\cite{wright1989crystalline,deGennes_Prost}, $A_0$ sets the scale of the bulk free energy
density, $\gamma$ controls the magnitude of the ordering,  $L$ is the elastic constant, $p=2\pi/q_0$
is the cholesteric pitch, $\epsilon_a$  is the (positive) dielectric anisotropy of the material and
$\epsilon_{\alpha\beta\gamma}$ is the Levi-Civita antisymmetric tensor with $\alpha$, $\beta$ and
$\gamma$ indexing the Cartesian components.

To match the properties of typical cholesterics, which have pitch $\approx  1\mu$m,
free energy density scale $\approx 10^5$Pa and elastic constant $\sim 6$ pN
we set the units of time and length to $1\mu$s and $\approx 0.03
 \mu$m respectively, and fix $A_0=1$, $\gamma=3$, $L=0.065$ and $q_0=2\pi/32$.
 The corresponding dimensionless chirality $\kappa=\sqrt{108L
 q_0^2/A_0\gamma}\approx0.3$ and reduced temperature
 $\tau=27(1-\gamma/3)/\gamma=0$~\cite{wright1989crystalline} are such that the
cholesteric phase is the lowest free energy phase at zero field~\cite{PhysRevA.28.1114,alexander2006stabilizing}.
 The degree of coupling with the latter is captured by the adimensional
 parameter ${\mathcal{E}}=\sqrt{27 \epsilon_a E_{\alpha}^2/(32\pi
 A_0\gamma)}$. Here we choose the values $\mathcal{E}_1=0.0377$,
 $\mathcal{E}_2=2\sqrt{2}\mathcal{E}_1$, $
\mathcal{E}_3=4 \mathcal{E}_1$, that span a range wide enough to straddle the bulk cholesteric-nematic
transition, see Supplementary Material (SM) for a numerical estimate of the onset of the transition~\cite{SM}.

The phenomenological equation of motion for ${\bf Q}$ is~\cite{ravnik2009landau}
\beq
\label{eq:4}
\dot{Q}_{\alpha \beta}= -\Gamma  \Bigl( \tfrac{\delta {\cal F}}{\delta Q_{\alpha \beta}} - \tfrac{1}{3} \text{Tr} \Bigl( \tfrac{\delta {\cal F}}{\delta \mathbf{Q}} \Bigr)  \delta_{\alpha\beta} \Bigr),
\eeq
\noindent where $\mathcal{F} = \int_V f$ is the total free energy and the collective isotropic rotational diffusion coefficient, $\Gamma$ is set equal to 0.5, corresponding to a rotational viscosity of $\approx 1$ Poise.
Eq.~\ref{eq:4} is solved numerically with a finite difference  scheme on a grid of size $N_x=N_y=N_z=96$ with periodic boundary conditions. The off-lattice positions of the centers of the spherical colloids (with radius $R_0=8$ corresponding to  $~0.25$ $\mu$m), ${\mathbf x}$, evolve according to an overdamped Langevin dynamics,
\bea
\label{eq:5}
\dot{x}_{\alpha}=F_\alpha/\xi+\sqrt{(2k_B T)/\xi}\,\zeta_\alpha(t),
\eea
\noindent where $\zeta_\alpha$ is a Gaussian white noise of unit variance, the
Stokes friction coefficient is $\xi=6\pi \eta R_0$ and the effective viscosity, $\eta$, is taken equal to the rotational one. ${\bf F}$ is the force acting on the
particle which comprises a steric and a liquid crystal mediated contribution (see SM~\cite{SM} and ~\cite{ravnik2009landau}).
The homeotropic anchoring at the colloid surface is enforced by setting
$Q_{\alpha\beta}=q (n_\alpha n_\beta-\delta_{\alpha,\beta}/3)$ where ${\mathbf n}$ is the unit vector normal to the surface
and $q$ is proportional to the magnitude of order ($q=1/3$ is appropriate for $\gamma=3$).
\begin{figure}[htb!]
   \centering
    \includegraphics[width=0.485\textwidth]{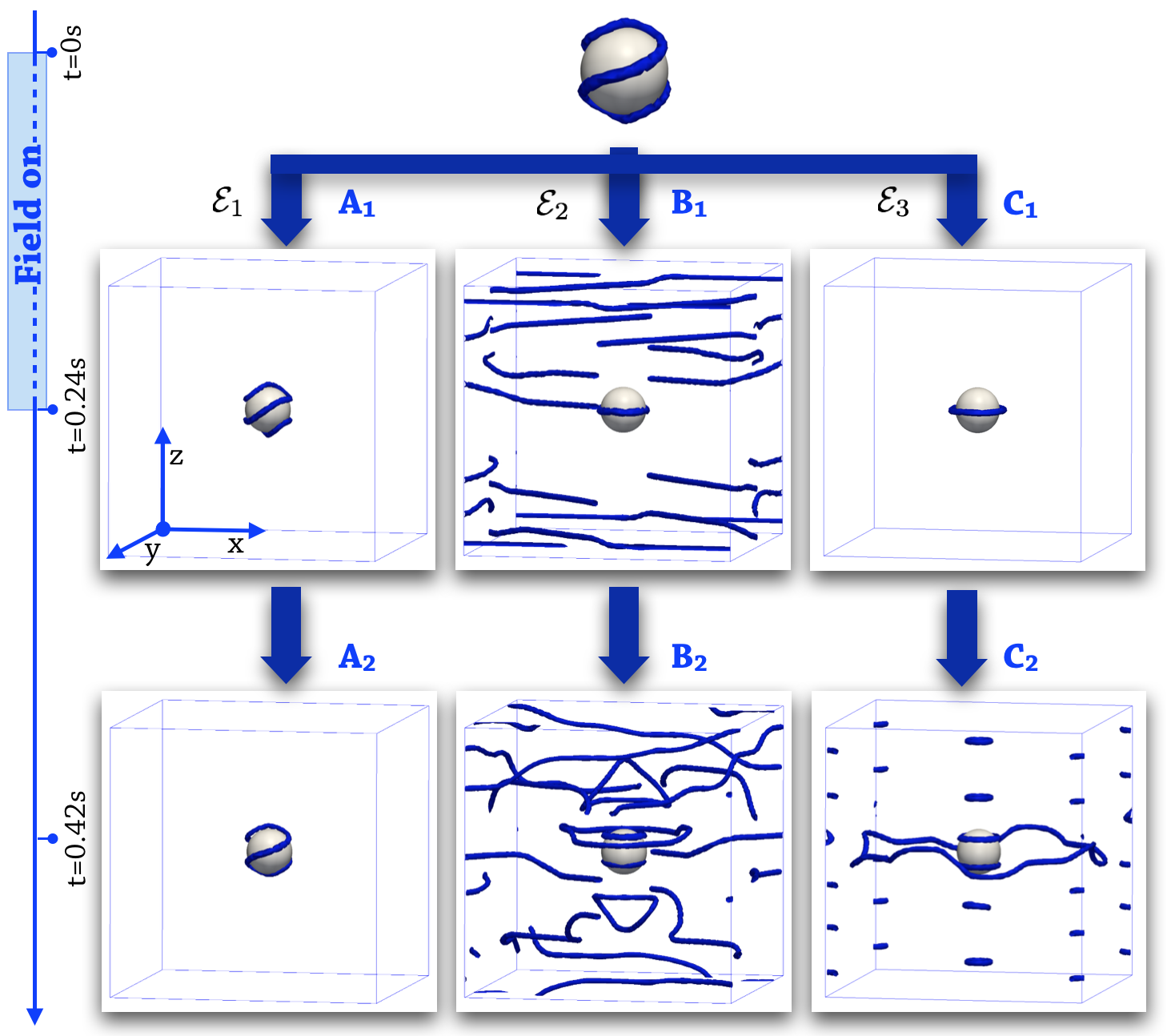}
     \caption{Evolution of an equilibrated cholesteric colloid (top) subject to an electric field pulse starting at $t=0$ and ending at $t=0.24$s. The asymptotic states at $t=0.24$s for the three increasing field strengths, $\mathcal{E}_1 < \mathcal{E}_2 < \mathcal{E}_3$ are shown in panels A$\rm{_1}$, B$\rm{_1}$ and C$\rm{_1}$, respectively. The corresponding post-pulse relaxed states at t=$0.42$s are shown in panels A$\rm{_2}$, B$\rm{_2}$ and  C$\rm{_2}$.}
 \label{Fig:fig1}
\end{figure}
As a first, reference case we discuss how switching on and off an electric
field affects the texture of a cholesteric hosting a single
colloid. The system was first relaxed in zero field in its
most stable state where the cholesteric  helix axis is along the $z$ direction
and the particle is embraced by a -1/2 twisted disclination
line~\cite{MarenduzzoCholColl-2010,jampani2011colloidal}. Next, an electric
field pulse along $z$ was applied from time $t=0$ to $t=0.24$s, see
Fig.~\ref{Fig:fig1}. The pulse duration is long enough to allow the system to
relax at all three field strengths  ${\cal E}_1$,  ${\cal E}_2$ and  ${\cal
E}_3$, see SM~\cite{SM}.
For each case, Fig.~\ref{Fig:fig1} illustrates the common initial condition,
the field-on relaxed configuration  and the post-pulse asymptotic state.
At the smallest field, $\mathcal{E}_1$, the field-relaxed configuration is
only slightly distorted (with a modest 20\% contour-length increase)
with respect to both the initial and post-pulsed
configurations, which are identical.  However, both the intermediate field
$\mathcal{E}_2$ and the largest one $\mathcal{E}_3$ are strong enough to induce a nematic
ordering of the medium so that the twisted disclination straightens up into
a Saturn ring defect in the field-relaxed configuration. Apart from this,
the system behaves differently at the two fields. First, the nematic order
induced by $\mathcal{E}_3$ is practically perfect, while it is only partial for
$\mathcal{E}_2$, where several {extended} disclination lines perpendicular to the field
direction are observed. Second, these disclinations reflect the presence of
several metastable states which render the relaxation dynamics at
$\mathcal{E}_2$ substantially slower than either at $\mathcal{E}_1$ or
$\mathcal{E}_3$, (see SM~\cite{SM}).

These differences become even more pronounced after the field pulse, when the
liquid crystal texture cannot relax back to the equilibrium cholesteric order because it
is trapped into metastable structures that are created during the post-pulse relaxation and
whose qualitative features depend on the pulse strength (i.e. the system has memory of its history).
In particular, after the strongest field ($\mathcal{E}_3$) is switched off, the
Saturn ring surrounding the colloid expands, while
two planar hoops normal to $\hat{\bf z}$ appear at the colloid surface close to the poles.
At the same time, a regular array of stacked disclination loops appears {above and below the colloid at a vertical spacing about $p/2$. These defects appear with the same relative colloid positioning in larger simulation cells, unlike the small disclinations on the vertical edges in panel C2 which, being finite-size effects, move further away from the colloid, see SM. Compared to $\mathcal{E}_3$,} the relaxation dynamics for $\mathcal{E}_2$ follows a qualitatively-different pathway.
While the defect evolution in proximity of the particle is similar to the $\mathcal{E}_3$ case, the network of
defects away from it is much more complex and arrests in an amorphous
structure. The post-pulse asymptotic free energies of the states subjected to the fields $\mathcal{E}_2$
and $\mathcal{E}_3$ exceed the initial one by respectively, $\approx 2.5\cdot10^4 k_BT$ and $\approx 2\cdot 10^4 k_BT$.
Such large offsets are indicative of  glassy-like landscapes where a plethora of energy barriers
much larger than thermal energy prevents relaxation to the lowest free-energy state.

As we discuss below, the rich morphology of the free energy landscape, which is both created and exposed by
the transient action of the external field, can be exploited to stabilize several metastable
states characterized by  different symmetries and different responses to an
external perturbation.  This can pave the way for the realization of switchable
devices based on cholesteric colloids, and controlled by time-dependent
external electric fields.

To illustrate this possibility, we considered an electric field modulated as a square wave with
strength $\mathcal{E}_3$ and switching time $\tau =0.02$s, that is longer then system relaxation
time, see SM~\cite{SM}.

\begin{figure}[htb!]
  \centering
    \includegraphics[width=0.485\textwidth]{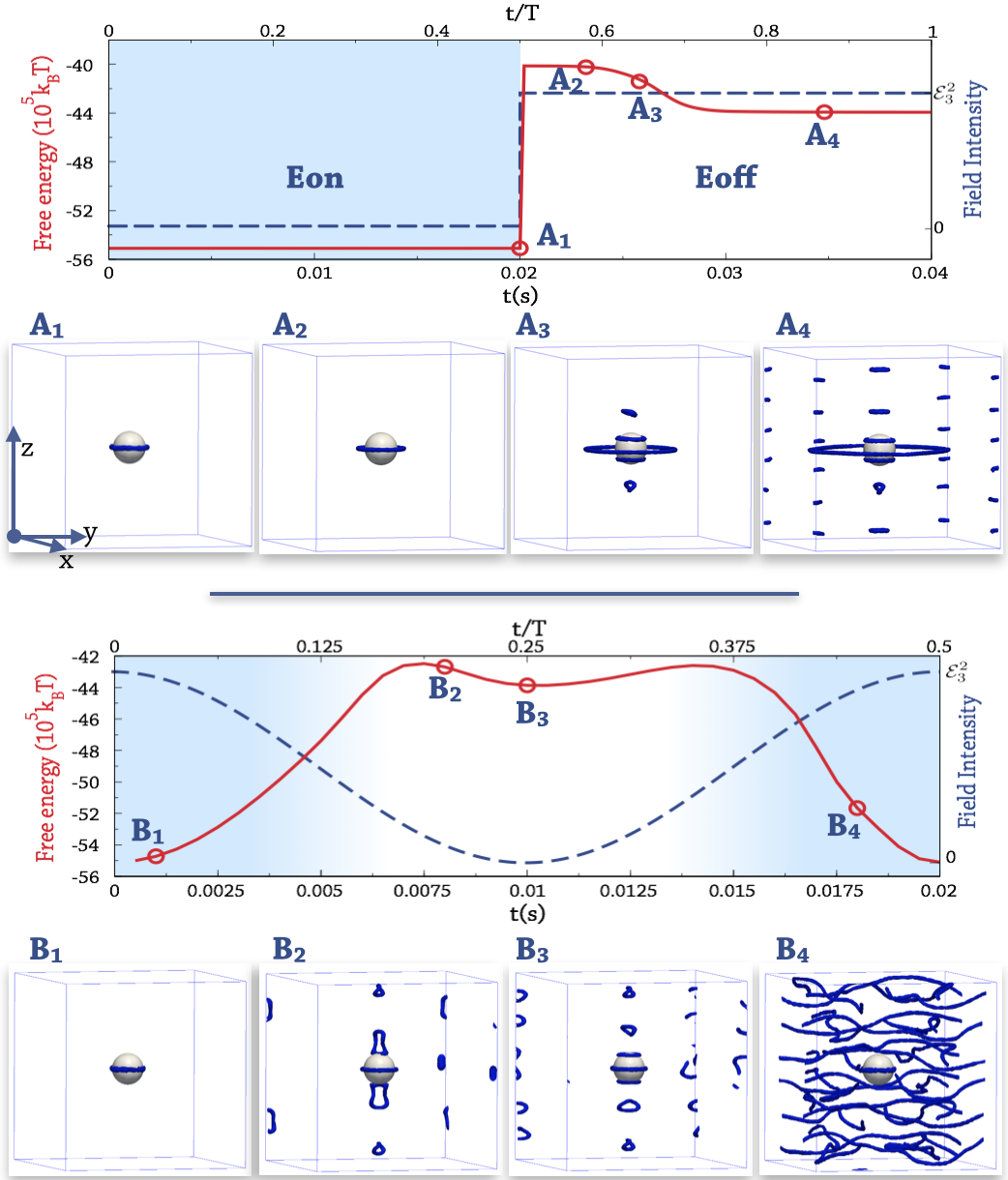}
       \caption{Steady-state evolution of a single cholesteric colloid in a periodically-modulated electric field. The cyclic time-evolution of the free energy and the field are shown in the first and third panel from the top for a square-wave and sine-wave modulation, respectively. For each case, four representative snapshots of the system dynamics are shown.}
       \label{Fig:fig2}
\end{figure}

Fig.~\ref{Fig:fig2} illustrates the time behavior of the system free energy of the
system over a period of the square-wave field. The superimposition
of the  free energy relaxation curves for different field switching events reveals
an almost perfect periodic steady-state response of the system which hops
between the two asymptotic field-on and field-off states shown in Fig.~\ref{Fig:fig1} in panels C$_1$ and C$_2$, see SM~\cite{SM}.
This periodic response is also observed in the dynamics of the defect
structures whose typical configurations within a single oscillation period are
shown in Fig.~\ref{Fig:fig2} panels A$_1$-A$_4$. Note that throughout the cycle, the
colloid position is practically fixed and the Saturn ring expands remaining planar.

Next we considered the steady-state system response to a sinusoidal wave.  Again, the colloid is
essentially motionless during the cycle and is surrounded by a single Saturn ring at the peak field
strength, when the medium consistently attains a purely-uniaxial state. Remarkably, however, the
sine-wave modulation leads to a significantly different evolution of defect patterns as the field
intensity decreases. In fact, the colloid surface becomes transiently encompassed by three hoops.
Furthermore, the periodic pattern of loops which bridge various cholesteric domains away from the
colloid now  grows  into an intricate defect network
before dissolving to leave a nematic state
when the field returns to its peak strength (see
Fig.~\ref{Fig:fig2} $B_4$).
The above phenomenology demonstrates that a time-dependent field provides a promising avenue to
reproducibly navigate a system of cholesteric colloids through its complex landscape of metastable states.
The remarkable underlying physical behaviour and its practical ramifications are aptly exposed by
extending considerations from a single colloid to a collection of particles.
In this case, cooperative effects are expected to arise from effective inter-colloid interactions mediated by
the intricately-evolving defect patterns~\cite{muvsevivc2006two,Tomar2012}.

\begin{figure}[htb!]
\includegraphics[width=0.485\textwidth]{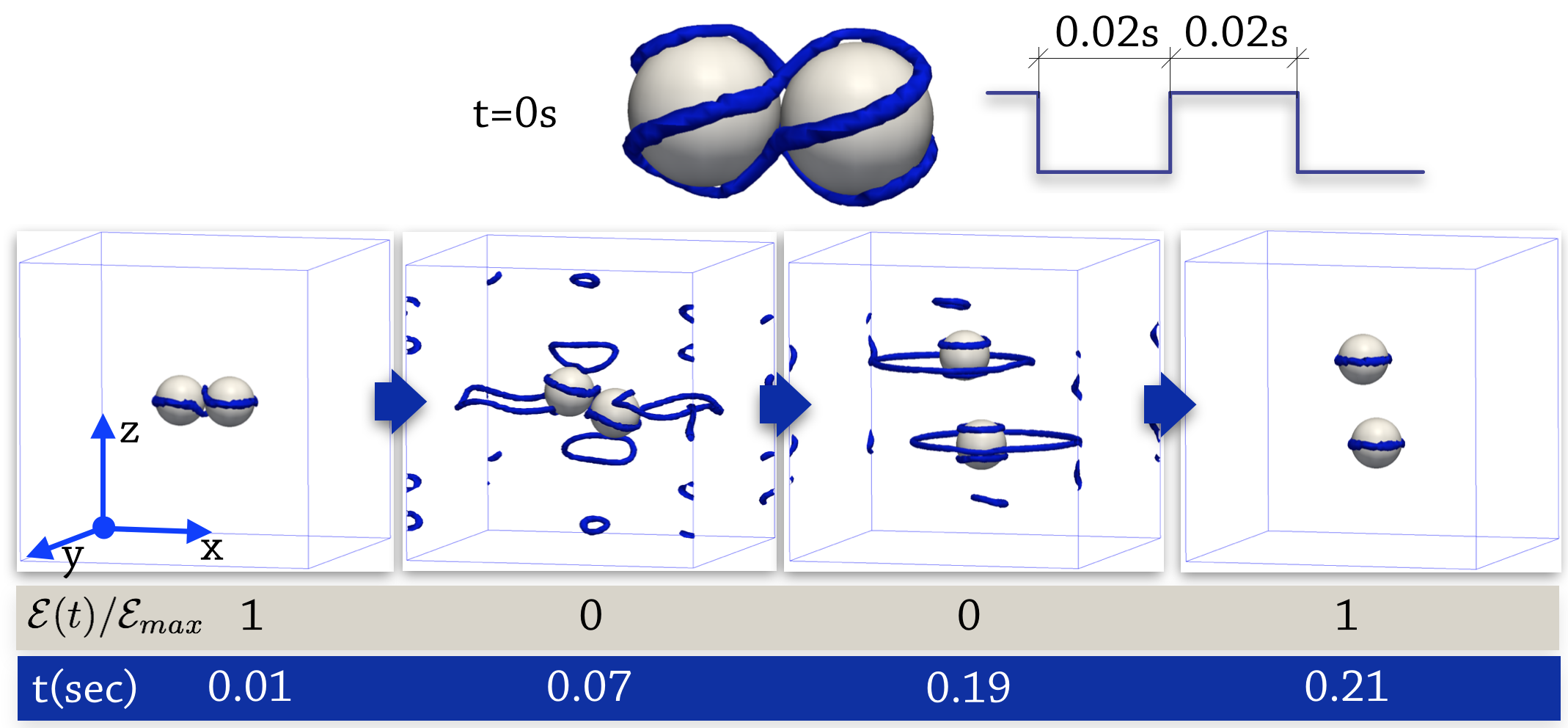}
\caption{An equilibrated  $\hat{\bf x}$-aligned cholesteric dimer is driven to the orthogonal, $\hat{\bf z}$-stacked configuration by a cyclic electric field modulated as a square-wave.}
\label{Fig:fig3}
\end{figure}

Accordingly, we considered a dimer oriented along the $x$ direction and relaxed in zero field. The
equilibrium state features two unlinked and unknotted disclination lines which wrap around the
particles (Fig.~\ref{Fig:fig3} and~\cite{,jampani2011colloidal,foffano2014dynamics}) thereby producing an
effective colloid-colloid  attraction of $\approx -4\cdot 10^2 k_B T$.  A square-wave electric field,
along $z$, with period $0.02 \text{s}$ and  strength $\mathcal{E}_3$ was next switched on.

The typical subsequent evolution of the system is illustrated in Fig.~\ref{Fig:fig3}.
When the field is switched on for the first time the
cholesteric undergoes a transition to the nematic state, associated
with the formation of a figure-of-eight entangled state around
the dimer~\cite{ravnik2007entangled,foffano2014dynamics}.
When the field is switched off, the system remains trapped in a metastable state, reminiscent of the one found
for a single inclusion (Fig.~\ref{Fig:fig1}, panel C$_2$). After this, however, the interaction between
the two colloids gives rise to a different and unexpected evolution.
In particular, after two cycles, the colloids displace, {mostly} along $\hat{\bf z}$ and,
when the field is switched on again, the figure-of-eight defect
is replaced by two unlinked Saturn rings.
{Thereafter the colloids are displaced by competing forces of about $\approx 30\text{pN}$ resulting from the topological restructuring of the disclination patterns following the field modulation, as detailed in SM. In particular, on one hand, the switching on of the field favours the approach of the colloids and their alignment perpendicular to the field. On the other hand, whenever the field is switched off the colloids are pushed further apart and towards a vertical alignment. At steady state, the balance of these competing effects leads the colloids to stack approximately on top of each other thus forming an  columnar arrangement along the $z$ direction, as shown in Fig.~\ref{Fig:fig3}.}
It should be emphasized that the observed colloidal reorientation and ordering along the field direction is a purely non-equilibrium effect driven by the defect rearrangement: accordingly, the free energy is higher for the columnar arrangement than for the figure-of-eight disclination.
Importantly, the {vertical colloid alignment is robustly attained upon varying the system size or by replacing the square-wave modulation with a sinusoidal one (see SM~\cite{SM})}.

\begin{figure}[tb!]
\includegraphics[width=0.485\textwidth]{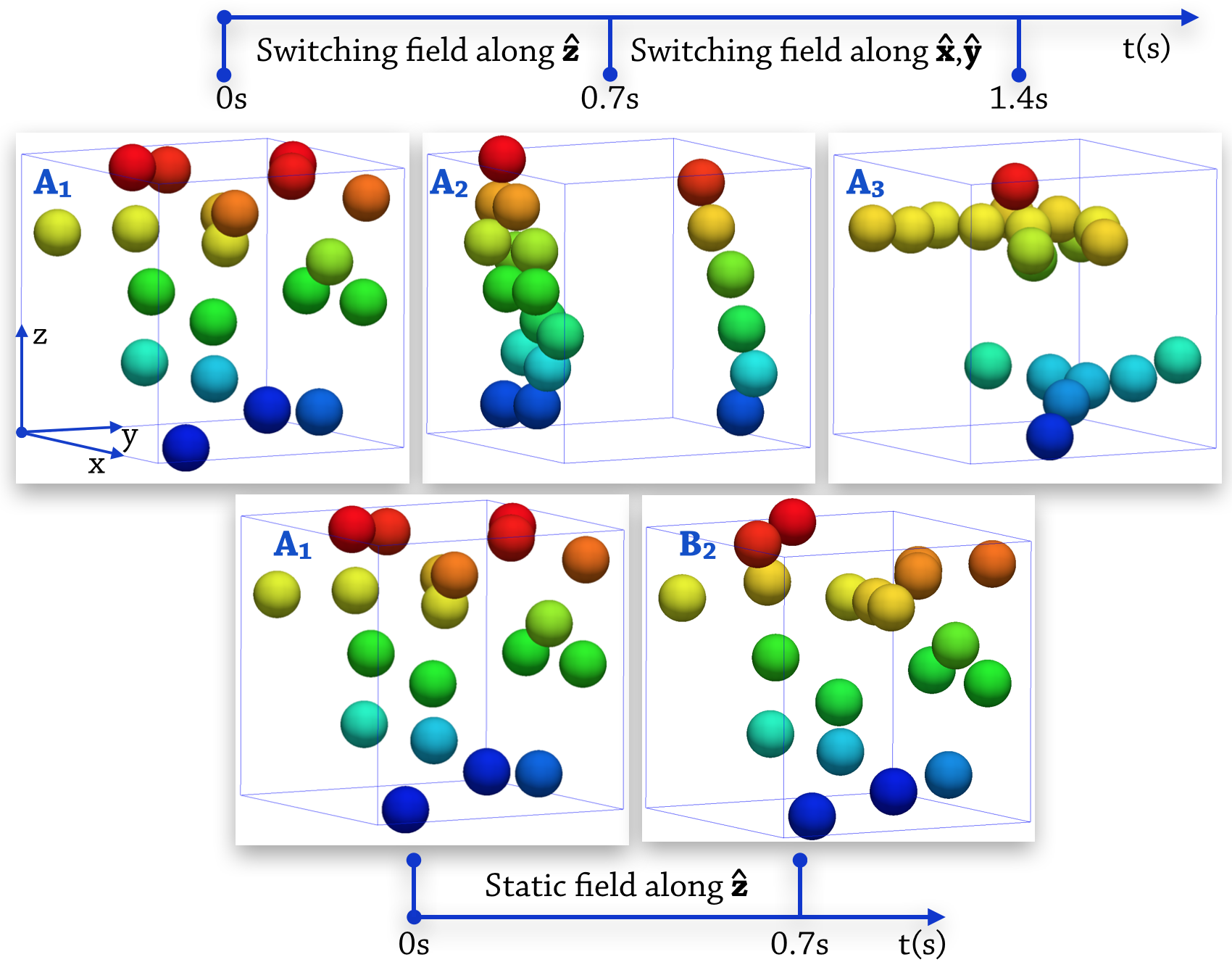}
\caption{Top Panel: Controlled columnar and planar alignment of several colloids with time-dependent electric fields.
An equilibrated random dispersion of colloids (A$_1$) is first subjected to a cyclic field along $\hat{\bf z}$, resulting in vertical columnar alignments (A$_2$). Next, the repeated on-off switching of the field in the $\hat{\bf x}$, and $\hat{\bf y}$ directions rearranges the colloids in horizontal planes (A$_3$). Bottom Panel: the same initial configuration (A$_1$) is only slightly perturbed (B$_1$) by the application of a static electric field in the  $\hat{\bf z}$ axis.}
\label{Fig:fig4a}
\end{figure}

We finally turn to the case of several interacting particles, which we address by considering a
colloid suspension of about $5\%$ in volume fraction. Starting from an equilibrated random initial
condition (A$_1$), we find that, as in the case of dimers, a time-dependent field (maximal strength
${\cal E}_3$, along the $z$ direction) drives a collective alignment and clustering of the particles
into colloidal columns along the field direction.  This non-equilibrium ordering is essentially
driven by the rearrangement of the disclination network created after field removal, that can
displace particles efficiently due to impulsive forces similar to those which we observed in the
case of a dimer. At subsequent switchings on of the field, the defect-mediated interactions induce
tighter columnar packing. This assembling and ordering process can be controlled by tuning not only
the field strength or frequency, but also its direction. For instance, starting from the state of
Fig.~\ref{Fig:fig4a} (A$_2$) and then cycling the field along $\hat{\bf x}$ and $\hat{\bf y}$, the
colloidal columns rearrange into more disordered aggregates, which mainly lie on the $xy$ plane
(Fig.~\ref{Fig:fig4a} (A$_3$)).

Importantly, the outcome is completely different when a static
field of same strength and direction is applied (see Fig.~\ref{Fig:fig4a} bottom panel). In this case, the kinetics entails a fast transition to the nematic state, where the colloids are accompanied by Saturn rings or
figure-of-eight defects, but we observe negligible particle motion, so that the colloidal
dispersion this time remains disordered.
In conclusion, we have studied the physics of colloidal dispersions in
cholesterics under a time-dependent electric field. For a single particle, we
showed that by tuning either the strength, or  the  time
modulation of the field, it is possible to reach a variety of metastable states,
characterised by distinctively different disclination networks. Colloidal
dimers and dispersions can  also be  repositioned under a time-dependent
field. Most remarkably, a time-dependent field can rearrange an initially
random dispersion into a set of field-aligned colloidal columns.
{This global
repositioning is possible because field-induced changes of the medium symmetry generate transient defects which, in turn, provide impulsive nonequilibrium forces favouring the axial alignment of neighbouring colloids.
We note that this effect is robust upon replacing the dynamical evolution of eq.~\ref{eq:4} with a more detailed one where hydrodynamic effects~\cite{PhysRevE.63.056702} are accounted for with a hybrid Lattice Boltzmann approach \cite{Desplat2001273}. As detailed in SM, we carried out this computationally-intensive analysis for the dimer case and observed the same two previously-described mechanisms which compete for repositioning of the colloids. In particular: at field on, colloids tend to be drawn closer and towards an alignment in the $xy$ plane while, at field off, they are pushed apart towards a $z$ alignment. The cyclic variations of distance and vertical alignment, which are appreciably larger with hydrodynamics than without, can be controlled by varying the interplay of these competing forces, e.g.~with the relative duration of the field on and off intervals, see SM.}
We hope that such rich phenomenology can stimulate further theoretical and experimental studies on colloid-liquid crystal composites under time-dependent fields, or other nonequilibrium perturbations.

We thank Oliver Henrich for helpful advice on the Ludwig simulation package. We acknowledge support from the Italian Ministry of Education grant PRIN No.~2010HXAW77.


\end{document}